# Automated Synthesis of Divide and Conquer Parallelism

Azadeh Farzan

University of Toronto

Victor Nicolet

Ecole Polytechnique

## Abstract

This paper focuses on automated synthesis of divide-and-conquer parallelism, which is a common parallel programming skeleton supported by many cross-platform multi-threaded libraries. The challenges of producing (manually or automatically) a correct divide-and-conquer parallel program from a given sequential code are two-fold: (1) assuming that individual worker threads execute a code identical to the sequential code, the programmer has to provide the extra code for dividing the tasks and combining the computation results, and (2) sometimes, the sequential code may not be usable as is, and may need to be modified by the programmer. We address both challenges in this paper. We present an automated synthesis technique for the case where no modifications to the sequential code are required, and we propose an algorithm for modifying the sequential code to make it suitable for parallelization when some modification is necessary. The paper presents theoretical results for when this *modification* is efficiently possible, and experimental evaluation of the technique and the quality of the produced parallel programs.

## 1. Introduction

Despite big advances in optimizing and parallelizing compilers, *correct and efficient* parallel code is often hand-crafted in a difficult and error-prone process. The introduction of libraries like Intel's TBB [39] was motivated by this difficulty and with the aim of making the task of writing parallel programs with good performance easy. These libraries offer efficient implementations for commonly used *parallel skeletons*. This makes it easier for the programmer to write code in the style of a given skeleton without having to make special considerations for important factors like scalability of memory allocation or task scheduling. Divide-and-conquer parallelism is one of the most commonly used of these skeletons.

Consider the function $sum$ that returns the sum of the elements of an array of integers. The code on the right is a sequential loop that computes this function.

```
sum = 0;
for (i = 0; i < |s|; i++) {
    sum = sum + s[i];
}
```

To compute the sum, in the style of divide and conquer parallelism, the computation should be structured as illustrated in Figure 1. The array $s$ of length $|s|$ is partitioned into $n + 1$ chunks, and $sum$ is individually computed for each chunk. Then the results of these partial $sum$ computations are joined (operator ⊙) into results for the combined chunks at each intermediate step, with the join at the root of the tree returning the result of $sum$ for the entire array. The burden of a correct design is to come up with the correct implementation of the join operator. Here, it is easy to quickly observe that the join has to simply return the sum of the two partial results. However, the correct join is not always as simple to formulate.

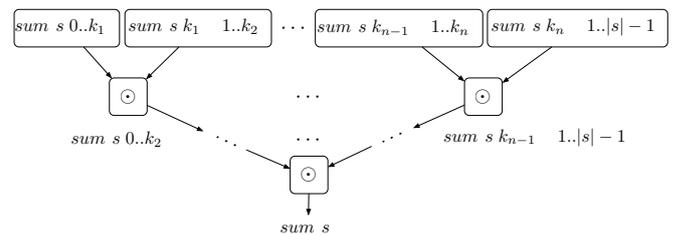

Figure 1: Divide and conquer style computation of $sum$.

Recent advances in program synthesis [2] demonstrate the power of synthesis in producing non-trivial computations. A natural question to ask is whether this power can be leveraged for this problem. In section 2, we present an example of a sequential loop that may be tricky to parallelize for an average programmer. However a *correct* join can be successfully synthesized, if the synthesis problem is framed properly.

In this paper, we focus on a class of divide-and-conquer parallel programs that operate on *sequences* (lists, arrays, or in general any collection with a linear iterator) in which the divide operator is assumed to be the default sequence *concatenation* operator (i.e. divide $s$ into $s_1$ and $s_2$ where $s = s_1 \bullet s_2$). In Section 4, we discuss how we use syntax-guided synthesis *efficiently* to synthesize these join operators. The main challenge for syntax-guided synthesis is to define a search space for synthesis that is expressive enough to include the desired solution, but limited enough to stay within the limits of scalability of current solvers. Moreover, in Section 7, we discuss how the proofs of correctness for all synthesized join operations can be automatically generated and checked. This addresses a second challenge that most synthesis approaches face, namely that the synthesized artifact is only guaranteed to be correct for the set of examples used in the synthesis process and not the entire (infinite) data domain of program inputs.

A general divide-and-conquer parallel solution is not always as simple as the diagram in Figure 1. Consider the function *is-sorted*($s$) which returns true if an array is sorted, and false otherwise. Providing the partial computation result, a boolean value in this case, from both sides to the join will not suffice. If both sub-arrays are sorted, the join cannot make a decision about the sortedness of the combined array. In other words, a join cannot be defined solely in terms of the sortedness of the subarrays.

It is clear to a human programmer that the join requires the last element of the first subarray and the first element of the second subarray to *connect the dots*. The extra information is as simple as remembering a value, but as we demonstrate with another example in Section 2, the extra information may need to be computed to be available to the join. Intuitively, this means that the worker threads (i.e. leaf nodes in Figure 1) have to be modified to compute this extra information so that a join operator exists. We call this modification of the code *an extension* for short. The necessity of the extension raises two questions: (1) does such extension always exist? and (2) can the overhead from the extension and the accompanying join overshadow the performance gains from parallelization? In Section 5, we answer both questions is the affirmative. The challenge for automation is then how to modify the original sequential loop so that it computes *enough* additional information so that (i) a join does exist, (ii) this join is efficient, and (iii) the overhead of the extension is not unreasonably high. Furthermore, we lay the theoretical foundations to answer this question, and in Section 6, we present an algorithm that produces a solution satisfying all aforementioned criteria.

In summary, in this paper, we make the following contributions:

- We present an algorithm to synthesize a join for divide-and-conquer parallelism when one exists. Moreover, these joins are accompanied by automatically generated machine-checked proofs of correctness (Sections 4, 7).
- We present an algorithm for automatic *extension* of non-parallelizable loops (when a join does not exist) to transform them into parallelizable ones (Section 6).
- We lay the theoretical foundations for when a loop is or can be made efficiently parallelizable (divide-and-conquer style), and explore when an efficient *extension* exists and when it can be automatically discovered (Sections 5, 6).
- We implemented our approach as a tool, PARSYNT, and present experimental results that demonstrate the efficacy of the approach and the efficiency of the produced solutions (Section 8).

## 2. Overview

In this section, we present an overview of our approach by the way of a few examples. We use successively more complicated examples to demonstrate the challenges of this problem, and how our approach deals with them. We refer to our linear collections as *sequences* that model any collection with a linear iterator.

*Second Smallest.* Consider the loop implementation of the function $min2$ that returns the second smallest element of a sequence, on the right. min and max are used for brevity, which can be replaced by their standard definitions $min(a,b) = a < b \,?\, a : b$ and $max(a,b) = a > b \,?\, a : b$. The correct join operator for a divide and conquer parallel implementation is
```
m = min(m_l, m_r);
m2 = min(min(m2_l, m2_r), max(m_l, m_r));
```
where the _l and _r suffixes distinguish the inputs to the join coming from *left* or *right*. It is easy to make the mistake of using the following join instead[1]:
```
m = min(m_l, m_r);
m2 = min(m2_l, m2_r);
```
which would return the incorrect result if partial results for the two sequences $[5, 1, 2, 7]$, where m is 1 and m2 is 2, and $[3, 8, 4, 9]$, where m is 3 and m2 is 4 were to be combined. Using a template which is based on the code of the loop body with unknowns (aka syntax-guided synthesis), we synthesize the correct join in 5s (more on this example in Section 4).

Figure 2: Second Smallest
```
m = MAX_INT;
m2 = MAX_INT;
for (i = 0; i < |s|; i++) {
    m2 = min(m2, max(m,s[i]));
    m = min(m, s[i]);
}
```

*Maximum Tail Sum.* Consider the function $mts$ that returns the maximum suffix sum of a sequence of integers (positive and negative). For example, $mts([1, -2, 3, -1, 3]) = 5$, which is associated to sum of the suffix $[3, -1, 3]$.

The code on the right illustrates a loop that implements $mts$. To parallelize this loop, the programmer
```
mts = 0;
for (i = 0; i < |s|; i++) {
    mts = max(mts + s[i], 0)
}
```
needs to come up with the correct join operator. In this case, even the most clever programmer would be at a loss, because there is no correct join. Consider the partial sequence $[1, 3]$, when extended by two different partial sequences $[-2, 5]$ and $[0, 5]$. We have:

$$mts([1,3]) = 4 \quad\quad mts([1,3]) = 4$$
$$mts([-2,5]) = 5 \quad\quad mts([0,5]) = 5$$
$$mts([1,3,-2,5]) = 7 \quad\quad mts([1,3,0,5]) = 9$$

The values of $mts$ for all partial sequences are the same, but the value of the $mts$ for the concatenation of sequences is different in the two cases. If a join *function* $\odot$ exists such that $mts(s \bullet t) = mts(s) \odot mts(t)$, then this function would have to produce two different outputs for $4 \odot 5$ for the two cases mentioned above. In other words, if the only information available to the join is the $mts$ value of the partial sequences, then the $mts$ value of the concatenation is not computable. What is to be done?

---
[1] In fact, a significant percentage of the undergraduate students in an algorithms class routinely make this mistake when given this exercise.

At this point, a clever programmer makes the observation that the loop is not *parallelizable* in its original form. In order to conclude that $mts([1, 3, -2, 5]) = 7$, beyond knowing that $mts([1, 3]) = 4$ and $mts([-2, 5]) = 5$, one needs to know that $sum([-2, 5]) = 3$.

Consider an extension of the previous sequential loop for $mts$ on the right. There is an additional loop variable, sum,

```
mts = 0;
sum = 0;
for (i = 0; i < |s|; i++) {
    mts = max(mts + s[i], 0);
    sum = sum + s[i];
}
```

that records the sum of all the sequence elements. In the sequential loop, it is a *redundant* computation. We call sum an *auxiliary* accumulator, for this reason. Now that the sum is available, the code can be parallelized with this join:

```
sum = sum_l + sum_r;
mts = max(mts_r, sum-r+mts_l);
```

With loops like this, the burden of the programmer is more than just coming up with the correct join code. She has to modify the original sequential code, by adding new computation, to make it parallelizable, and then devise the correct join for the new sequential loop. The algorithm presented in Section 5 does exactly that. We provide an overview of our algorithm by illustrating how it works for the $mts$ example to *discover* the *auxiliary* information about the sum.

Consider the general case of computing $mts(s)$ sequentially, where $|s| = n$. Imagine the point in the middle of the computation, when the sequence $s$ has been processed up to and including index $i$. We have the following sequence of *recurrence*-like equations that represent the unfolding of this sequential loop starting from index $i + 1$:

$$mts_{i+1} = max(mts_i + s[i+1], 0) \quad (1)$$

$$mts_{i+2} = max(mts_{i+1} + s[i+2], 0) \quad (2)$$

$$\ldots$$

When the (unfolded) computation above starts, the initial value of $mts_i$ is not available to it, since it is being computed in parallel by a different thread simultaneously. The challenge is how to (mostly) do the computation above, without having the value of $mts_i$ in the beginning, and *adjust* the computed value accordingly once the value of $mts_i$ becomes available. Let us start by examining Equation (1):

$$mts_{i+1} = max(mts_i + s[i+1], 0)$$
$$= max(max(mts_i + s[i+1], s[i+1]), 0)$$
$$= max(mts_i + s[i+1], max(s[i+1], 0))$$
$$= max(mts_i + s[i+1], max(s[i+1], 0))$$
$$= max(mts_i + s[i+1], mts([s[i+1]])$$

In the last line, the second term (under $max$) is computing the $mts$ as if the element $s[i + 1]$ is the first element of the (input) sequence. So, one can imagine that $mts_i$ (i.e. $mts(s[0..i])$) is computed by one thread, while a second thread (so far) computes $mts([s[i + 1]])$ (the remaining part of the sequence seen so far). To compute the value of $mts(s[0..i + 1])$, the join code needs the aforementioned partially computed values **and** the extra value $s[i + 1]$. Our algorithm (presented in Section 6), at this point *conjectures* $s[i + 1]$ as an extra auxiliary information and '+' as its accumulator, since it is actively looking for *accumulators* to *learn*. Moving to Equation (2) (skipping the simplification steps):
$mts_{i+2} = max(mts_{i+1} + s[i + 2], 0) = ...$
$= max(mts_i + s[i + 1] + s[i + 2], mts([s[i + 1..i + 2]))$
following the same line of argument as above, our algorithm discovers that it needs to calculate and pass along $s[i + 1] + s[i + 2]$. At this point, the algorithm realizes that the *conjectured* accumulator from the last round makes this information available. Therefore, not having seen anything new, it stops and concludes that the auxiliary accumulator $sum = sum + s[k]$ (where $k$ is the current iteration number) is the new auxiliary computation required to make the loop parallelizable.

## 3. Background and Notation

This section introduces the notation used in the remainder of the paper. It includes definitions of some concepts that are new, and some that are already known in the literature.

### 3.1 Sequences and Functions

We assume a generic type $Sc$ that refers to any of the scalar types used in programs, such as `int`, `float`, `char`, and `bool`, when the specific type is not important in the context. The significance of the type is that scalars are assumed to be of *constant* size, and therefore, any constant-size representable data type is considered scalar in this paper. Moreover, from the point of view of time complexity, we assume all operations on scalars to be constant-time.

Type $\mathcal{S}$ defines the set of all *sequences* of elements of the type $Sc$. For any sequence $x$, we use $|x|$ to denote the length of the sequence. $x[i]$ (for $0 \leq i < |x|$) denotes the element of the sequence at index $i$, and $x[i..j]$ denotes the subsequence between indexes $i$ and $j$ (inclusive). Concatenation operator $\bullet : \mathcal{S} \times \mathcal{S} \to \mathcal{S}$ is defined over sequences in the standard way, and is associative. . This sequence type stands in for *arrays*, *lists*, or any other linear collections that admit iterators and an *associative* composition operator (e.g. concatenation)

**Definition 3.1.** *A function $h : \mathcal{S} \to D$ is called rightwards iff there exists a binary operator $\oplus : D \times Sc \to D$ such that for all $x \in \mathcal{S}$ and $a \in Sc$, we have $h(x \bullet [a]) = h(x) \oplus a$*

Note that the notion of associativity for $\oplus$ is not well-defined, since it is not binary operation defined over a set.

$$\begin{aligned}
e \in \mathsf{Exp} ::=\ & e \bigcirc e' & e, e' \in \mathsf{Exp} \\
| & x & x \in \mathsf{Var} \\
| & s[e] & s \in \mathsf{SeVar}, e \in \mathsf{Exp} \\
| & k & k \in \mathbb{Z} \\
| & \text{if } be \text{ then } e \text{ else } e'
\end{aligned}$$

$$\begin{aligned}
\mathsf{Program} ::=\ & c; c' & c, c' \in \mathsf{Program} \\
| & x := e & x \in \mathsf{Var}, e \in \mathsf{Exp} \\
| & b := be & b \in \mathsf{BVar}, be \in \mathsf{Exp} \\
| & \mathtt{if}(e)\{c\}\mathtt{else}\{c'\} & be \in \mathsf{Exp}, c, c' \in \mathsf{Program} \\
| & \mathtt{for}(i \in D)\{c\} & i \in \mathsf{Iterator}
\end{aligned}$$

Figure 3: Program Syntax. The binary $\bigcirc$ operator represents any arithmetic operation $(+, -, *, /)$, $\oslash$ operator represents any comparator $(<, \leq, >, \geq, =, \neq)$. $D$ is an iteration domain.

**Definition 3.2.** *A function $h : \mathcal{S} \to D$ is called leftwards iff there exists a binary operator $\otimes : Sc \times D \to D$ such that for all $x \in \mathcal{S}$ and $a \in Sc$, we have $h([a] \bullet x) = a \otimes h(x)$.*

Associativity of $\otimes$ is not well-defined as well.

**Definition 3.3** (Single-Pass Computable). *Function $h$ is single-pass computable iff it is rightwards or leftwards.*

In this paper, our focus is on single-pass computable functions on sequences (of scalars). The assumption is that their implementation (leftwards or rightwards) is given in the form of an imperative sequential loop.

### 3.2 Homomorphisms

Homomorphisms are a well-studied class of mathematical functions. In this paper, we focus on a special class of homomorphisms, where the source structure is a set of sequences with the standard concatenation operator.

**Definition 3.4.** *A function $h : \mathcal{S} \to D$ is called $\odot$-homomorphic for binary operator $\odot : D \times D \to D$ iff for all sequences $x, y \in \mathcal{S}$ we have $h(x \bullet y) = h(x) \odot h(y)$.*

Note that, even though it is not explicitly stated in the definition above, $\odot$ is necessarily associative on $D$ since concatenation is associative (over sequences). Moreover, $h([\,])$ (where $[\,]$ is the empty sequence) is the unit of $\odot$, because $[\,]$ is the unit of concatenation. If $\odot$ has no unit, that means $h([\,])$ is undefined. For example, function $head(x)$, that returns the first element of a sequence, is not defined for an empty list. $head(x)$ is $\circledast$-homomorphic, where $a \circledast b = a$ (for all $a, b$) but $\circledast$ does not have a left unit element.

### 3.3 Programs and Loops

For the presentation of the results in this paper, we assume that our sequential program is written in a simple imperative language with basic constructs for branching and looping. We assume that the language includes scalar types `int` and `bool`, and a collection type `seq`.

The syntax of our input programs is presented in Figure 3. We forego a semantic definition since it is intuitively clear. For readability we use a simple iterator, and integer index (instead of the generic $i \in D$), and use the standard array random access $a[i]$ to refer to the next element. In principle, any collection with an iterator and a split function would work. There has been a lot of research on iteration spaces and iterators (e.g. [52] in the context of translation validation and [23] in the context of partitioning), that formalize complex traversals by abstract iterators. Programs with nested loops are admitted, however our tool always focuses on parallelizing one inner loop at a time.

*State and Input Variables.* every variable that appears on the lefthand side of an assignment statement in a loop body is called a *state variable*, and the set of state variables is denoted by SVar. Every other variable is an *input variable*, and the set of input variables is denoted by IVar. The sequence that is being read by the loop is considered an *input variable*, since it is only read and not modified. Note that $\mathsf{SVar} \cap \mathsf{IVar} = \varnothing$.

### Model of a Loop Body

We introduce a formal model for the body of a loop, which has no other loops nested inside it. The loop body consists of assignment and conditional statements only.

Let $\mathsf{SVar} = \{s_1, \ldots, s_n\}$. The body is modelled by a system of (ordered) recurrence equations $E = \langle E_1, \ldots E_n \rangle$, where each equation $E_i$ is of the form $s_i = \exp_i(\mathsf{SVar}, \mathsf{IVar})$.

**Remark 3.5.** *Every non-nested loop in the program model in Figure 3 can be modelled by a system of recurrence equations (as defined above).*

This can be achieved through a translation of the loop body from the imperative form through the functional form. In Appendix A, we include a description of how this can be done, but also refer the interested reader to [3, 24] for a more complete explanation. We provide an example that captures the essence of this translation, which is through transformation of conditional statements into conditional expressions.

**Example 3.6.** Consider the long version of the second smallest example from Section 2 on the right, where the auxiliary functions *min* and `max` are not used

```
m = MAX_INT;
m2 = MAX_INT;
for (i = 0; i < |s|; i++) {
  if m > s[i] then
     if m2 > m then m2 = m ;
  else
     if m2 > s[i] then m2 = s[i];
  if m > s[i] then m = s[i];
}
```

anymore, and instead the code complies with the syntax in Figure 3.

The loop body is then modelled by recurrence equations on the right, through the transformation of the conditional statements above into the assignment statements that use conditional expressions (if $be$ then $e$ else $e'$).

```
m2 = if m2 < (if m > s[i] then m else s[i])
        then m2
        else (if m > s[i] then m else s[i]);
m = if m < s[i] then m else s[i];
```

# 4. Synthesis of Join

The premise of this section is that the sequential loop under consideration is parallelizable, in the sense that a join operation for divide-and-conquer parallelism exists. The loop body is represented by a system of recurrence equations $E$ in the style of Section 3.3 throughout this section. All formal and algorithmic developments in the this paper use $E$ as a model of the loop body, and therefore, we use the term "loop body" to refer to $E$ whenever appropriate.

## 4.1 Parallelizable Loops

We start by formally defining when a loop is parallelizable. We define a function $f_e$ that models $E$, so that we can link the computation defined by $E$ to homomorphisms. Let $E = \langle s_1 = \exp_1(\mathsf{SVar}, \mathsf{IVar}), \ldots, s_n = \exp_n(\mathsf{SVar}, \mathsf{IVar}) \rangle$ be the system of recurrence equations that represents the body of a loop. Let $\mathsf{IVar} = \{\sigma, \iota, i_1, \ldots, i_k\}$ where $\sigma$ is the sequence variable, and $\iota$ is the current index (which we assume to be an integer for simplicity) in the sequence, and $\vec{i} = \langle i_1, \ldots, i_k \rangle$ captures the rest of the input variables in the loop body. Let $\mathcal{I}$ be the set of all such $k$-ary vectors.

For a loop body $E$, define function $f_E = f_1 \bowtie f_2 \bowtie \ldots \bowtie f_n$, where each $f_i : \mathcal{S} \times int \times \mathcal{I} \to type(s_i)$ is a function such that $f_i(\sigma, \iota, \vec{i})$ returns the value of $s_i$ at iteration $\iota$ with input values $\sigma$ and $\vec{i}$. It is, by definition, straightforward to see that $\langle f_1(\sigma, \iota, \vec{i}), \ldots, f_n(\sigma, \iota, \vec{i}) \rangle$ represents the state of the loop at iteration $\iota$.

**Definition 4.1.** *A loop, with body $E$, is parallelizable iff $f_E$ is $\odot$-homomorphic for some $\odot$.*

Definition 4.1 basically takes the encoding of the loop as a *tupled* function, and declares the loop parallelizable if this function is homomorphic. It is important to note that *parallelizable* here is not used in the broadest sense of the term. It is limited to divide-and-conquer parallelism.

## 4.2 Syntax-Guided Synthesis of Join

We use syntax-guided synthesis (SyGus)[2] to synthesize the join operator for parallelizable loops. The goal of program synthesis is to automatically synthesize a correct program based on a given specification. In SyGus, syntactic constraints are used to make the synthesis problem more tractable. In this section, we describe what our specification is, and what syntactic constrains are used. Beyond that, this paper is not concerned with how SyGus works. One (out of many available) SyGus solvers is used to solve instances that are fully defined with these two parameters.

First, in this paper we focus on synthesizing *binary* join operators. The restriction is superficial, in the sense that any other statically fixed number would work. And, it is not hard to imagine an easy generalization to a parametric join.

**Correctness Specification**

The homomorphism definition 3.4 provides us with the correctness specification to synthesize a join operator $\odot$ for a loop body $E$ (or rather $f_E$ to be precise). It is important to note that the SyGus solver requires a bounded set of possible inputs, and therefore, the correctness specification is formulated on symbolic inputs of bounded length $K$. A join $\odot$ is a solution of the synthesis problem if for all sequences $x, y$ of length less than $K$, we have $f_E(x \bullet y) = f_E(x) \odot f_E(y)$.

There is a tension between having a small enough $K$ for the solver to scale, and having a large enough $K$ for the answer to be correct for inputs that are larger than $K$. We use small enough values for the solver to scale, and take care of general correctness by automatically generating proofs of correctness (see Section 7).

**Syntactic Restrictions**

The syntactic restrictions are defined through a pair of a *sketch* and a grammar for expressions. A *sketch* is a partial program containing *holes* (i.e. unknown values) to be completed by expressions in the state space defined by the grammar. The sketch is an ordered set of equations $E = \langle s_1 = \exp_1(\mathsf{SVar}, \mathsf{IVar}), \ldots, s_n = \exp_n(\mathsf{SVar}, \mathsf{IVar}) \rangle$. Intuitively speaking, it is produced from the loop body by replacing occurrences of variables and constants by holes. Note that the input to a join is the result produced from two worker threads, to which we refer to as *left* and *right* threads. To contain the state space of solutions described by this sketch, we distinguish two different types of holes.

*Left holes* $??_L$ are holes that can be completed by an expression over variables from both the left and the right threads. *Right holes* $??_R$ will be filled with expressions over variables from the right thread only. We define a compilation function $\mathcal{C}$ as

$$\begin{aligned}
\mathcal{C}(k) &= ??_R \\
\mathcal{C}(x) &= \begin{cases} ??_R & \text{if } x \in \mathsf{IVar} \\ ??_L & \text{if } x \in \mathsf{SVar} \end{cases} \\
\mathcal{C}(x[e]) &= ??_R \\
\mathcal{C}(op(e_1, .., e_n)) &= op(\mathcal{C}(e_1), ..., \mathcal{C}(e_n))
\end{aligned}$$

where $e$ is an expression, $op$ is an operator from the input language, $x$ is a variable, and $k$ is a constant. The sketch for the join code will then be

$$\mathcal{C}(E) = \langle s_1 = \mathcal{C}(\exp_1(\mathsf{SVar}, \mathsf{IVar})), \ldots, s_n = \mathcal{C}(\exp_n(\mathsf{SVar}, \mathsf{IVar})) \rangle$$

where each hole in $\mathcal{C}(\exp_i)$ ($1 \le i \le n$) can be completed by expressions in a predefined grammar.

The grammar for replacing of the unknowns is similar to the general expression grammar described in Figure 3, i.e. expressions over a predefined set of operators, a set of variables, and integer/boolean constants. The state space considered is bounded by considering expressions of bounded depth from this set.

**Example 4.2.** Consider the second small-
```
m2 = min(??_L, max(??_L, ??_R));
m  = min(??_L, ??_R);
```
est example from Section 2. The sketch (for the code in Figure 2) is illustrated on the right. It is clear that the join presented in Section 2 can be simple discovered using this Sketch when the unknowns are replaced by appropriate single variables.

### 4.3 Efficacy of Join Synthesis

SyGuS has its limitations: (i) the syntactic limitations imposed for scalability run the risk of leaving a correct candidate out of the search space, and (ii) most SyGuS solvers use bounded verification to ensure that the candidate satisfies the correctness specification and therefore the synthesized solution may not be correct in the general case. We discuss how to deal with (ii) in Section 7. To characterize the impact of the former problem, we will provide formal conditions under which the approach presented in Section 4.2 is successful in synthesizing a correct join.

**Definition 4.3.** *Function $g$ is a weak-right inverse of function $f$ iff $f \circ g \circ f = f$.*

Intuitively, $g$ produces a sequence $s'$ (out of a result $r = f(s)$) such that $f(s') = r$. In [33], this very notion of a weak-right inverse was used to produce parallel implementations of functions, given that $s'$ is always a bounded list independent of $|s|$. Our join synthesis is also guaranteed to succeed in these cases.

**Proposition 4.4.** *If the loop body $E$ is parallelizable and the weak-right inverse of $f_E$ returns a list of bounded length, then there is a $\odot$ in the space described by the sketch of $E$ such that $f_E$ is $\odot$-homomorphic.*

Proposition 4.4 (see Appendix B for a proof sketch) provides us with the guarantee that under the conditions given, the state space defined for the join does not miss an existing solution. It, however, only defines a subset of functions for which, the join synthesis succeeds. There are many (simple) functions whose weak right inverse is not bounded, where a join is successfully synthesized. For example, function $length$ is a simple example, where the weak right inverse always has to be a list of the same length as the original list (which would lead to an inefficient parallel implementation).

In principle, without any syntactic restrictions, any constant-time join can be synthesized when the entire right-hand side of each equation in the join is one unknown. However, synthesis will not scale in this case. In Section 6.1, we extend the scope of efficacy of join beyond Proposition 4.4, while maintaining the scalability of synthesis.

## 5. Parallelizability

The $mts$ (maximum tail sum) example from Section 2 demonstrates that a loop is not always parallelizable. Here (and later in Section 6), we discuss how a loop can be *extended* to become parallelizable.

### 5.1 Homomorphisms and Parallelism

It has been well-known that there is a strong connection between homomorphisms and parallelism [16, 18, 33]. Specifically that, homomorphic functions admit efficient parallel computation.

**Theorem 5.1** (First Homomorphism Lemma [9]). *A function $h : S \to Sc$ is a homomorphism if and only if $h$ can be written as a composition of a map and reduce operation.*

The *if* part of Theorem 5.1 basically states that a homomorphism has an efficient parallel implementation, since efficient parallel implementations for *maps* and *reductions* are known. The *only if* part is equally important, because it indicates that a large class of *simple and reasonable* parallel implementation of a list function ought to be homomorphisms; this includes, every implementation that can fit TBB's generic skeleton.

Theorem 5.1 poses two important research questions: (1) How can one check if a sequential loop is an implementation of a homomorphic function over sequences? (2) If it is not, can we modify the sequential code so that it becomes implementation of a homomorphic function? The algorithm in Section 6 addresses both questions. Here, we build the formal foundations that provide the context for the algorithm and the specific design choices for it.

***Non-homomorphic Functions.*** There is a simple observation, which was made in [19] (and probably also in earlier work), that any non-homomorphic function can be made homomorphic in a rather trivial way:

**Proposition 5.1.** *Given a function $f : S \to Sc$, the function $f \bowtie \iota$ is $\diamondsuit$-homomorphic where*

$$\iota(x) = x \quad \wedge \quad (a, x) \diamondsuit (b, y) = f(x \bullet y)$$

Operation $\bowtie$ denotes *tupling* of the two functions in the standard sense $f \bowtie g(x) = (f(x), g(x))$. Basically, the $\iota$ remembers a copy of the string that is being processed and the partial computation results $f(x)$ and $f(y)$ are discarded by $\diamondsuit$, which then computes $f(x \bullet y)$ from scratch. It is clear that a parallel implementation based on this idea would be far less efficient that the sequential implementation of $f$. Therefore, not every homomorphic extension is a good choice for parallelism. Next, we identify a subset of homomorphic extensions that are computationally efficient.

### 5.2 Computationally Efficient Homomorphisms

Let us assume that $f : S \to D$ is a *single-pass linear time computable* (see Definition 3.3) function and elements of $D$ are tuples of scalars. The assumption that the function is linear time comes from the fact that we target loops without any loops nested inside, which (by definition) are linear-time computable (on the size of their input sequence). In other words, for any sequence $s$, $f(s)$ can be computed in time linear on $|s|$. Note that $|s|$ is the only variable in the complexity argument below, and everything else is considered a constant.

Consider the (rightwards) sequential computation of $f$ that is defined using the binary operator $\otimes$ as follows:

$$f(y \bullet a) = f(y) \otimes a$$

where $a \in Sc$, $f : \mathcal{S} \rightarrow D$, and $\otimes : D \times Sc \rightarrow D$. The fact that $f$ is single-pass linear time computable implies that the time complexity of computing $f$ at every step (of the recursion above) is constant-time.

Consider an extension of $f$, that is also $\odot$-homomorphic (for some $\odot$), which we call $f'$.

**Proposition 5.2.** *The (balanced) divide-and-conquer implementation based on $f'$ has the same asymptotic complexity as $f$ (i.e. linear time) iff the asymptotic complexity of $\odot$ is sub-linear (i.e $o(n)$).*

*Proof.* See Appendix C for a simple proof. □

The simplest case is when $\odot$ is constant time. We call these *constant homomorphisms* for short. In this paper, (i) we provided a synthesis method that exclusively synthesizes *constant time* join operators, and (ii) in the next section we focus on discovery of constant-size auxiliary variable, namely scalars, which will guarantee the inputs to the join are constant-size and therefore the join is constant-time.

Let us briefly consider the super-constant yet sub-linear case for joins, to clarify why this paper only focuses on the constant-time joins. Based on the definition of homomorphism, we know:

$$f'([x_1, \ldots x_n]) = f'([x_1, \ldots x_k]) \odot f'([x_{k+1}, \ldots x_n])$$

For $\odot$ to be super-constant in $n$, $f'([x_1, \ldots x_k])$ (respectively, $f'([x_{k+1}, \ldots x_n])$) has to produce a super-constant output, that is then processed by $\odot$. Constant-size inputs cannot yield super-constant time executions. Since the output of $f$ was assumed to be constant-sized (scalars are by definition constant sized), the output to $f'$ is super-constant only due to the *auxiliary* information, required to have a homomorphism, but not required for the sequential computation of $f$. We believe the automatic discovery of auxiliary information of this nature to be a fundamentally hard problem. The sub-linear auxiliary information is often the result of a clever algorithmic trick that improves over the trivial linear auxiliary (i.e. remembering the entire sequence as was discussed in Proposition 5.1). The field of efficient *data streaming algorithms* [1, 4] includes examples of such clever algorithms. In conclusion, despite the fact that the join synthesis can be adapted to handle such examples, since we consider the auxiliary information synthesis in this case to be always necessary and extremely difficult to do automatically, we do not target this class for automation in this paper.

## 6. Synthesizing Parallelizable Code

Let us assume that the synthesis of join from Section 4.2 has failed. Then we know that either (1) the function is not a homomorphism, or (2) the function is a homomorphism, but syntactic restrictions for the synthesis of join were too restricted to include the correct operator. We deal with case (1) here, and case (2) in Section 6.1.

**The Algorithm**

In Section 2, we provided a step by step explanation of how considering the unfoldings of the computation of $mts$ would lead to discovery of the *auxiliary accumulator sum*. Here, we present the algorithm that discovers these accumulators by unfolding the loop body starting from an arbitrary state. This arbitrary state symbolically represents the point at which a sequential computation is meant to be divided into two parallel computations, and the algorithm is intuitively following in the footsteps of the second computation.

---

**Algorithm 1:** Homomorphic Extension Computation

**Data**: A set of recurrence equations $E$
**Data**: A constant cost bound $C$
**Result**: A set of recurrence equations $E'$
**Function** *Extend(E)*
  $E' \leftarrow \emptyset$;
  **for** *each $s_i = \exp_i$ in $E$ (in order)* **do**
    **if** *Solve("$s_i = \exp_i''$") = null* **then**
      report failure
    **else**
      $E' \leftarrow E' \cup Solve("s_i = \exp_i''")$;
  return $E'$;
**Function** *Solve(s = Exp(SVar, IVar))*
  Initially $k = 1$, Aux $= \emptyset$, $\sigma_0 = \langle s_0^0, \ldots, s_n^0 \rangle$
  **while** Aux $\neq$ OldAux **do**
    OldAux $\leftarrow$ Aux ;
    $\tau \leftarrow$ unfold$(\sigma_0, s, E, k)$;
    $\tau \leftarrow$ *Normalize*$(\tau, E, C)$;
    **if** $\tau = null$ **then**
      return $null$;
    $\mathcal{E} \leftarrow$ collect$(\tau, \text{SVar})$;
    **for** *each $e$ in $\mathcal{E}$* **do**
      **if** *e already covered by something in* Aux
      **then**
        Add the accumulator and continue
      **else**
        Create a new variable in Aux, with expression $e$.
    $k \leftarrow k + 1$;
  Aux $\leftarrow$ *remove-redundancies*(Aux);
  return Aux;
**Function** *Normalize*($\exp_1, \exp_2, E, C$)
  **Output**: An expression $e = \exp_1$ with a subexpression $\exp_2$ and Cost$_{\text{SVar}}(e) \leq C$
  "Intentionally left with no implementation."

---

Algorithm 1 illustrates our main algorithm. The algorithm starts from the first equation in the loop body $E$,

and examines each recurrence equation in order (by calling *Solve()*).

In function *Solve()*, we start from an arbitrary state $\langle s_0^0, \ldots, s_n^0 \rangle$. The while loop iteratively unfolds the expression in the style of equations (1-2) in Section 2. unfold simply computes the $k$-th unfolding. *normalize* is the key step in this algorithm. It takes the unfolded expression and returns an equivalent expression of the form $\exp(e_1, \ldots, e_m, e)$ where $e$ only refers to the input variables and each $e_i$ is an expression which includes one occurrence of a state variable from SVar at depth 1. The intuition is that the 'remainder' of each $e_i$ (besides the one occurrence of the state variable) needs to captured by an auxiliary accumulator. For example in the $mts$ example from Section 2, in both unfoldings we had $m = 1$, and $e_1 = mts(\ldots) + sum(\ldots)$. The simplification steps that we did perform for $mts_{i+1}$ and $mts_{i+2}$ are exactly the steps that are performed by *normalize*.

*collect* gathers these $e_i$'s in $\mathcal{E}$. For each $e_i$, we check if it can already be accounted for using existing auxiliary accumulators in Aux. Otherwise, a new auxiliary variable and accumulator is recorded.

*normalize* relies on a cost function to ensure that all instances of the state variables appear at a depth lower than a predetermined constant ($C$) in the normal form.

**Definition 6.1.** *The* cost function $\text{Cost}_V : \exp \to \text{Int} \times \text{Int}$ *assigns to any expression $e$, a pair $(d, n)$ where $d$ is $Max_{v \in V} \, dep_e(v)$ where $dep_e(v)$ is the depth of the maximum occurrence of $v$ in $e$, and $n$ is $\sum_{v \in V} occ_e(v)$, where $occ_e(v)$ is the number of occurrences $v$ in $e$.*

*normalize can* always succeeds when the appropriate auxiliary accumulator exist, as long as all the algebraic rules that are needed for simplification are at its disposal. In fact, in the algorithm presented in Figure 1, we assume that this idealized version of *normalize*. This is why no particular implementation is presented. This implies that if *normalize* returns $null$ then no good expression exists. The existence of an efficient implementation (to replace this idealized version) depends on the state space of the equivalent expressions.

The process of normalization can be formalized by a rewrite system $R$ and a set of equations $A$ such that $R$ is convergent modulo $A$. Given a set of algebraic rules $R$ and a set of equations $A$, we define the *cost minimizing* normalization procedure as the successive application of rewrite rules in $R$ and substitutions using equations in $A$ such that we reduce an expression to an equivalent expression with minimal cost. In our implementation, we use simple heuristic for normalization with algebraic rules like *associativity*, *commutativity*, and *distributivity* with each rule applied only in the direction that it would reduce the cost of the expression, which we will evaluate in Section 8. However, many other heuristics for normalization can be imagined [28, 29, 35].

The following Proposition states the correctness of Algorithm 1. Note that in all the formal statements about Algorithm 1, we assume the idealized version of *normalize* as discussed.

**Proposition 6.2.** *If Algorithm 1 terminates successfully and reports no new auxiliary variables, then the loop body corresponds to a $\odot$-homomorphic function (for some $\odot$). If it terminates successfully and discovers new auxiliary accumulators, then the loop body extended with them corresponds to a $\odot$-homomorphic function (for some $\odot$).*

### 6.1 Completeness

Algorithm 1 is correct when it terminates successfully. However, it can terminate reporting a failure. Here, we formally discuss when Algorithm 1 succeeds, and when it fails.

**Existence of Constant Homomorphic Extensions**

First, independently of any algorithm, we deliberate on the question of existence of an efficient solution. A constant homomorphic extension includes *constantly many constant-size auxiliary variables* (see Section 5.2). Remember that extending the loop to a homomorphism with a single linear-size auxiliary variable (Proposition 5.1) is always possible.

**Theorem 6.1.** *If $f$ is leftwards and rightwards linear time, then there exists a binary operator $\odot$ and a $\odot$-homomorphic extension of $f$ that is a constant homomorphism.*

*Proof.* The proof of this theorem is rather involved and is included in Appendix D. $\square$

Theorem 6.1 defers the existence of a constant homomorphic extension to the fact that function $f$ can be computed in single-pass linear-time both leftwards and rightwards on a sequence. This may not always be true. Even if a function is sing-pass linear-time computable in one direction, it may be more expensive to compute in the other direction. In [38], the problem of language recognition of a string that is divided between two agents is discussed, and a complexity argument is provided that a restriction to a single-pass (in a certain direction) could increase the time complexity exponentially. Here, we provide a simple example to pass the intuition on to the reader.

**Remark 6.3.** *Let $h_n : \{0, 1\}^* \to bool$ be defined as $h_n(w) = \text{true}$ iff $w[n] = 1$ (i.e. the $n$-th letter). $h$ is rightwards but not leftwards linear time computable in $n$.*

The argument for this claim is included in Appendix G. For a function like $h$, there does not exist a *constant* homomorphic extension.

**Proposition 6.4.** *If a function $h$ is not leftwards (resp. rightwards) linear-time computable, then there exists no homomorphic extension of $h$ that is linear-time computable.*

This is a simple consequence of the *Third Homomorphism Theorem* [17]. A homomorphism naturally induces a leftwards and a rightwards function. Existence of a constant homomorphic extension would imply that the function should be single-pass linear-time computable leftwards and rightwards which is a contradiction.

**Corollary 6.5.** *There exists a linear time loop for which there does not exist a linear time homomorphic extension.*

This corollary is significant, because it implies that not every simple sequential loop (i.e. one that models a single-pass linear-time computable function) does necessarily have a constant homomorphic extension.

**Completeness of the Algorithm**

Since not every single-pass linearly computable function can be extended into a constant homomorphism, we can conclude that there exists no algorithm that is complete for the entire class of these function. It remains to determine, for those that can be extended to constant homomorphisms (see Theorem 6.1), how 'complete' Algorithm 1 is.

**Theorem 6.2.** *For a function $f$, if there exists a constant $\odot$-homomorphic extension $f \bowtie g$, then there exists a $\circledast$-homomorphic extension where*

$$(f(x), g(x)) \circledast (f(y), g(y)) =$$
$$(exp(f(x), f(y), g(y)), exp'(f(x), g(x), f(y), g(y)))$$

*that is, the value of $f(x \bullet y)$ component of the join does not depend on $g(x)$.*

*Proof.* The proof is included in Appendix E. □

Theorem 6.2 is very significant. It is the key in the completeness argument for Algorithm 1. The way the algorithm operates is that it tries to discover $exp(f(x), f(y), g(y))$ through the unfoldings of $f(x \bullet y)$, and does not have access to $g(x)$. To make any claims about completeness of Algorithm 1, one has to argue that it is sufficient to look at $f(x)$, $f(y)$, and $g(y)$ (and not $g(x)$).

**Theorem 6.3** (Completeness). *If there exists a constant homomorphic extension of the loop body $E$, then there is a finite set of algebraic rules $\mathcal{R}$, and a run of Algorithm 1 where the algorithm succeeds in discovering this extension.*

Proof of Theorem 6.3 appears in Appendix F. Note that existence of a *constant homomorphic* extension was formulated in Theorem 6.1. As stated by Theorem 6.3, the completeness result holds under the assumption of availability of a sufficient set of algebraic rules, and for the nondeterministic algorithm that theoretically may explore all equivalent expressions. Any implementation will be determinized using a particular efficient heuristic for this exploration. In Section 8, we evaluate our heuristic using a set of benchmarks.

Finally, there is an important observation that Algorithm 1 includes information that can be helpful to join synthesis. Imagine join synthesis fails, while a constant-time join exists. When Algorithm 1 terminates successfully and declares that no new auxiliary accumulators were discovered, then it becomes clear that the problem lies with the join; i.e the state space of the synthesis devised for the join is too restricted. In this instance, the *normalized* expression used for the discovery of the accumulators contains hints about the shape of the join operator, and with these new hints, join synthesis can be re-instantiated to succeed.

## 7. Correctness of the Synthesized Programs

As noted in Section 4, the synthesized join operators are not guaranteed to be correct for all input sequences. We use Dafny [26] program verifier to prove the correctness of the join operator, and therefore the synthesized parallel code on all inputs. Dafny is an *interactive* program verifier, in which users have to provide the necessary invariants (or hints) for the proofs. Automated parallelization is meant to reduce the programmer's burden in correctly devising parallel programs. It is therefore unreasonable to assume that the programmer would invest the extra effort of guiding a program verifier to establish a proof. It is essential that the proofs of correctness are also automatically generated with the code. That is exactly what we do.

The proof of correctness of a join operator for a function (over a sequence) is an induction argument over the length of the sequence. The generality of this induction, in the sense that it is independent of the specific function, enables us to devise a generic proof template that we can instantiate for each function and its corresponding join operator.

We use an example to illustrate the generic induction template. Let us assume that we want to verify the correctness of the join operator synthesized to parallelize the $mts$ loop body. As mentioned in Section 3.3, as part of the task of synthesizing the join, we obtain a functional variant of the loop body. We use this functional variant to model the loop body as a collection of functions in Dafny. The general rule is that every state variable in the loop body is modelled by a different function. Figures 4(a) and 4(b) illustrate the functions that model state variables `sum` and `mts` from Section 2 respectively. The Auxiliary function `Max` also needs to be separately defined in Dafny, which appears in Figure 4(c).

The next step is to model the synthesized join. Each state variable is modelled by a distinct join function. Therefore, in this case, we introduce a function to model the join for the `Sum` function as illustrated in Figure 4(d), and another one for the `Mts` function as illustrated in 4(e).

Once all components of the divide and conquer program are modelled, we are ready to state the correctness of the join as two Dafny lemmas, one for each join operator. Let us start with `Sum`. The lemma in Figure 4(f) states correctness of join for `Sum`. This lemma is basically stating that `Sum` is `SumJoin`-homomorphic as stated by definition 3.4, through its `ensures` clause. Dafny cannot prove this lemma automatically. It can prove the lemma, however, if it is given a hint about how to formulate an induction argument. One needs to specify what the base case is, and how to break the argument to prove the induction step.

```
function Sum(s: seq<int>): int        function Mps(s: seq<int>): int              function Max(x: int, y: int): int
{ if s == [] then 0 else              { if s == [] then 0 else                    { if x < y then y else x }
    Sum(s[..|s|-1]) + s[|s|-1] }          Max(Mts(s[..|s|-1]) + s[|s|-1], 0)}

function SumJoin(sum_l: int, sum_r: int): int    function MtsJoin(mts_l: int, sum_l: int, mts_r: int, sum_r: int): int
{ sum_l + sum_r }                                { Max(mts_r, mts_l + sum_r) }

lemma HomomorphismSum(s: seq<int>, t: seq<int>)       lemma HomomorphismMts(s: seq<int>, t: seq<int>)
  ensures Sum(s + t) == SumJoin(Sum(s), Sum(t))         ensures Mts(s + t) == MtsJoin(Mts(s), Sum(s), Mts(t), Sum(t))
{                                                     {
  if t == [] { assert(s + t == s); }                    if t == [] { assert(s + [] == s); }
  else {                                                else {
     calc {                                                calc { Mts(s + t);
        Sum(s + t);                                           == {HomomorphismSum(s,t[..|t|-1]);
        == {assert (s + t[..|t|-1]) + [t[|t|-1]] == s + t;}        assert (s + t[..|t|-1]) + [t[|t|-1]] == s + t;}
        SumJoin(Sum(s), Sum(t)); }                            MtsJoin(Mts(s), Sum(s), Mts(t), Sum(t)); }
  }                                                     }
}                                                     }
```

Figure 4: The encodings of *sum* (a) *mts* (b), *max* (c). The encoding of join in two parts for *sum* (d) and *mts* (e). The proofs that the joins form a homomorphism in two parts for *sum* (f) and *mts* (g). `s + t` denotes the concatenation of sequences `s` and `t` in Dafny, `[]` is the empty sequence while `|s|` stands for the length of the sequence `s`. The lemmas are proved by induction through the separation of the base case `t == []` and the induction step. The `calc` environment provides the hints necessary for the induction step proof.

There are several ways, for a human, to formulate the proof of correctness of `HomomorphismSum` lemma in Dafny. The unique aspect of the illustrated proof (among many others) is that it does not employ any hints that are specific to function `Sum`. The following hints are provided in 4(f):

- The proof is an induction argument on the length of `t` (the second sequence), with the base case of an empty sequence (reflected in the `if` statement condition).

- In the base case, Dafny should be aware of the fact that an *empty* `t` basically simplifies the reasoning about `s + t` to reasoning about `s`.

- In the induction step (i.e. the `else` part), the guide required to prove the two sides equivalent is the hint that the last element of `t` should be peeled off, and the induction hypothesis needs to be recalled on `s + t[..|t|-1]` (i.e. concatenation of `s` and `t` minus its last element).

These hints are generic enough to be applicable to any other function. Let us illustrate that by presenting the proof for the correctness of `MtsJoin`. Note that `Mts` calls `Sum` in its definition, and therefore the proof of correctness of the join operator for `Mts` has to call on the proof of the correctness of the join for `Sum` (namely, the `HomomorphismSum` lemma). Figure 4(g) illustrates this proof. The proof is identical to the 4(f), with the minor difference of recalling of the already proved lemma in 4(f).

The general idea of generating Dafny proofs follows the idea presented using the `Mts` example above. The general template of the proof is the same as the one for `HomomorphismSum` or `HomomorphismMts` lemmas. However, the proof for each state variable $v$ in addition recalls (as hints) the proofs (of smaller induction instances) of all state variables $u$ that are mentioned on the right-hand side of the assignment to $v$ in the code of the synthesized join. The proof for each instance can be discharged based on $E$ (the model of the loop body) and the synthesized join operator.

It is important to be clear about limitations of automated proofs. The scalar data types used in these implementations must belong to the Dafny language. Automation in provers, in general, is limited by the existence of decision procedures for a subset of all data types (used in common programming languages) and operations on them. These inherent limitation in lack of decision procedures carry over to limitations on the reach of our automated proof generator.

## 8. Experiments

Our parallelization technique is implemented in a prototype tool called PARSYNT, for which we report experimental results in this section.

### 8.1 Implementation

We use CIL [36] to parse C programs, do basic program analysis, and locate inner loops. The loop bodies are then converted into functional form, from which a sketch and the correctness specification is generated. ROSETTE [47] is used as our backend solver, with a custom grammar for synthesizable expressions. In addition to the narrowing of the search space by using *left* and RIGHT holes (unknowns) as discussed in Section 4, we use type information to reduce the state space of the search for the solver. We also bound the size of the synthesizable expressions, especially when the loop body contains *non-linear operators*, which are difficult for solvers to handle. To sidestep this problem, we produce a more constrained join sketch with a smaller search space for when non-linear operators are involved.

### 8.2 Evaluation

**Benchmarks.** We collected a set of benchmarks to evaluate the effectiveness of our approach. Table 1 includes a complete list. The benchmarks are all in C and form a

| | sum | min | max | average | hamming | length | 2nd-min | mps | mts | mss | mts-p | mps-p | poly | is-sorted | atoi | dropwhile | balanced-() | 0*1* | count-1's | line-sight | 0after1 | max-block-1 |
|---|---|---|---|---|---|---|---|---|---|---|---|---|---|---|---|---|---|---|---|---|---|---|
| Aux required? | no | no | no | no | no | no | no | yes | yes | yes | yes | yes | no | yes | yes | yes | yes | yes | yes | yes | yes | yes |
| Join Synt time? | 1.6 | 2.0 | 1.7 | 22.9 | 1.5 | 1.6 | 6.1 | 3.1 | 3.9 | 29.4 | 82.2 | 77.1 | 115.2 | 7.0 | 84.4 | 3.7 | 19.8 | 1.4 | 3.0 | 6.6 | 7.6 | 7.5 |
| #Aux required | – | – | – | – | – | – | – | 1 | 2 | 1 | – | – | 1 | 1 | 1 | 1 | 2 | 1 | 1 | 1 | 1* |

Table 1: Experimental results for performance of PARSYNT over all benchmarks. Times are in seconds. "–" indicates that no relevant data can be reported in this case. *: tool succeeds in finding 1 out 2 necessary auxiliaries. Auxiliary synthesis and proof generation/checking times negligible in all cases. Hardware: laptop with 8G RAM and Intel dual core m3-6Y30.

diverse group of functions:

- `min`, `max`, `length`, `sum`, `is-sorted` and `average` are standard functions over lists. `2nd-min` is the second smallest example discussed in Section 2.

- `mps` (maximum prefix sum), `mts` (maximum tail sum), and `mss` (maximum segment sum) are programming pearls from [33, 43]. `mps-p` and `mts-p` are the variations ( from [43]) where in addition to the value of the corresponding sum, the position that defines it is also returned.

- `poly` computes the value of a polynomial at a given point when the coefficients are given in a list. `atoi` is the standard (string to integer) function from C.

- `balanced-()` checks if a string of brackets (()) is balanced. `count-1's` counts the number of blocks (i.e. contiguous sequence) of 1's in a sequence of 0's and 1's. `max-block-1` returns the length of the maximum block of (contiguous) 1's. `0after1` is a language recognizer that accepts strings in which a 0 has been seen after a 1. $0^*1^*$ is a regular expression filter. `hamming` computes the hamming distance between two strings.

- `dropwhile` (from Haskell) is a *filter* that removes from the beginning of a sequence all the elements that do not satisfy a given predicate. `line-sight` (from [33]), determines if a building is visible from a source in a line of other buildings of various height.

*Performance of* PARSYNT  Table 1 lists the times spent in join synthesis. The times for auxiliary accumulator synthesis are negligible (about 1-2ms on average). Note that the subset of the benchmarks that were parallelizable (in original form), no attempt to discover auxiliaries was made (indicated by "–" in the table). When auxiliary variables were needed, Table 1 reports how many were discovered. With one exception (i.e. `max-block-1`), the auxiliary synthesis always succeeds. In this case, 1 of the 2 auxiliaries required is discovered, but the tool fails to discover the second one. Manual inspection convinced us that the discovery of the second one would be possible if the basic set of algebraic identities were enriched with more sophisticated indentities. Finally, the time to generate and check proofs is negligible compared to the join synthesis times, with most cases taking about 1s, and the most expensive case taking about 7s.

*Quality of the Synthesized Code*  We manually inspected all synthesized programs, and as programmers, we cannot produce a better version for any of them other than $0^*1^*$. For this one, only one of the two auxiliary accumulators discovered was strictly necessary, and the other was redundant. There is, however, a negligible performance difference between this version and the optimized one. Here, we evaluate the performance of the synthesized programs. The quality of a parallel implementation, depends on many parameters beyond the algorithmic design (which is the topic of study in this paper). We use Intel's Thread Building Blocks (TBB) [39] as the library to implement the divide-and-conquer parallel solutions that we synthesize. TBB is a popular runtime C++ library, that offers improved performance scalability by dynamically redistributing parallel tasks across available processors, and accommodates portability across different platforms. It supports divide-and-conquer parallelism, so transforming our solutions (from ROSETTE) into a TBB-based implementation became a trivial task. To evaluate the quality of the generated parallel code, we used Proliant DL980 G7 with 8 eight-core Intel X6550 processors (64 cores total) and 256G of RAM running 64-bit Ubuntu.

Figure 5 illustrates the speedups of our parallel solutions over the input sequential programs. The input size is about 2bn and the grain size is set at 50k. Note that for those benchmarks that auxiliary computation was required for parallelization, the parallel version is more expensive per iteration than the original sequential input. It is clear that the speedups are linear on the number of cores up to around 32 cores. A study [13] of TBB's performance has shown that scaling well above 32 cores is a known problem with TBB. It is due to TBB's scheduling overhead and not due to a design problem in our produced parallel programs. We separately measured the overhead of TBB, by limiting the number of cores to 1. The slowdown (over sequential) was on average negligible; the average slowdown close to 1, with a standard deviation of 0.04.

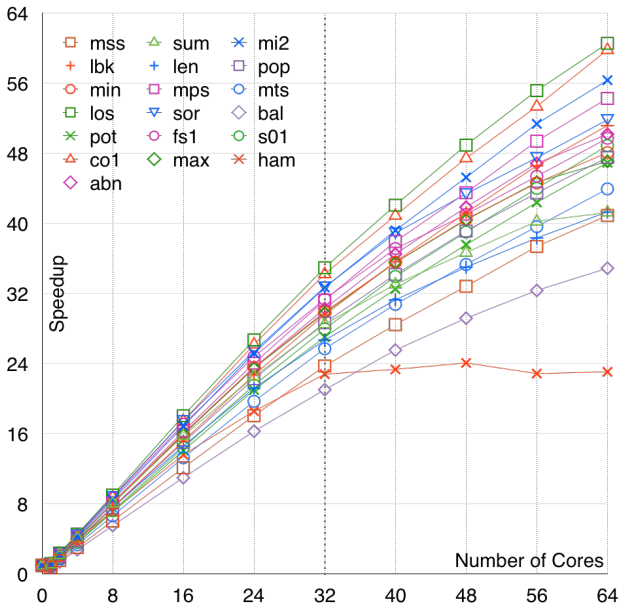

Figure 5: Speedups relative to the sequential implementation

## 9. Related Work

*Parallelizing Compilers and Runtime Environments*
Automatic parallelization in compilers is a prolific field of research, with source-to-source compilers using complex methods to parallelize generic code [5, 11, 37, 51] or more specialized nested loops with polyhedral optimization [6, 7, 48]. Another body of work specific to reductions and parallel-prefix computations [10, 20, 25] deals with problems when the dependencies cannot be broken, but parallelization is still possible. Another approach to automatic parallelization is to break the static dependencies at runtime, Galois[40] being a good example of this category. Handling irregular reductions, when the operations in the loop body are not immediately associative, has been explored by employing techniques such as data replication or synchronization [21]. We do not have enough space to do the mountain of research on parallelizing compilers justice.

Our approach in this paper is not based on the idea of correct source-to-source transformation. Rather, the aim is to use search, in the style of synthesis, in a space that includes many incorrect programs. But once the correct program is found, it does not rely on a runtime environment support. It is correct on all platforms. Without this degree of freedom, it would be impossible to discover parallel implementations of many of our benchmarks.

More recently in [42], symbolic execution is used to identify and break dependences in loops that are hard to parallelize. Since we produce correct parallel implementations, we do not have to incur the extra cost of symbolic execution at runtime. That said, the scope of applicability of [42] and our approach are not comparable. In the related area of distributed computation, there has been research on producing MapReduce programs automatically, for example by using specific rewrite rules [41], or based on program semantics [27], or synthesis [44].

*Homomorphisms and Parallelism* The closest category of papers to ours are the ones that use homomorphisms for parallelization. There has been previous attempts in using the derivation of list homomorphisms for parallelization, such as methods based on the third homomorphism theorem [16, 18], those based on function composition [15], their less expressive/more practical variant based on matrix multiplication [43], methods based on the quantifier elimination [31] as well as those based on recurrence equations [8]. We will discuss the most closely related one here.

One recent work [33] uses the *third homomorphism theorem* and the construction of the *weak right inverse* to derive parallel programs. The approach in [33] requires much more information from the programmer compared to our approach. The programmer needs to provide the leftwards **and** rightwards sequential implementations of a function to get the parallel one. This is often unreasonable, specially when the parallel version has a twist, since coming up with the leftwards implementation of a function could be as complex (and time consuming) as parallelizing it in the first place. Intuitively, by providing the reverse computation, the programmer is providing the information that we compute automatically here in Section 6. By contrast, we only need one (reference) sequential implementation as input. Moreover, this method works well when the size of the weak right inverse is constant (i.e. it does not depend on the length of the list). In examples as simple as the $length$ function, this does not hold. In this paper we go beyond, and easily synthesize solutions for functions like $length$.

The theoretical aspect of [33] was later extended to trees in [32], and generalized for lists in [30]. But these papers do not provide a practical way of generating implementations. In [12], a study of how functions may be extended to homomorphisms is presented, but no algorithm is provided.

*Synthesis and Concurrency* Synthesis techniques have been leveraged for the parallel programs before. Instances include synthesis of distributed map/reduce programs from input/output examples [44], optimization and parallelization of stencils [22, 45], synchronization synthesis [50], concurrent data structures synthesis [46], concurrency bug repair [49]. Other than use of synthesis, these problem areas and the solutions have very little in common with this paper. The most relevant and recent work is SYMPA [14] which uses synthesis to parallelize a reference sequential implementation by analyzing data dependencies. Our work entirely subsumes this work, since the prefix information used in [14] is only a special case of our auxiliary accumulators.

# References

[1] N. Alon, Y. Matias, and M. Szegedy. The space complexity of approximating the frequency moments. In *Proceedings of the Twenty-eighth Annual ACM Symposium on Theory of Computing*, STOC '96, pages 20–29, 1996. ISBN 0-89791-785-5.

[2] R. Alur, R. Bodík, E. Dallal, D. Fisman, P. Garg, G. Juniwal, H. Kress-Gazit, P. Madhusudan, M. M. K. Martin, M. Raghothaman, S. Saha, S. A. Seshia, R. Singh, A. Solar-Lezama, E. Torlak, and A. Udupa. Syntax-guided synthesis. In *Dependable Software Systems Engineering*, pages 1–25. 2015.

[3] A. W. Appel. Ssa is functional programming. *SIGPLAN Not.*, 33(4):17–20, Apr. 1998. ISSN 0362-1340.

[4] B. Babcock, S. Babu, M. Datar, R. Motwani, and J. Widom. Models and issues in data stream systems. In *Proceedings of the Twenty-first ACM SIGMOD-SIGACT-SIGART Symposium on Principles of Database Systems*, PODS '02, pages 1–16, 2002. ISBN 1-58113-507-6.

[5] D. F. Bacon, S. L. Graham, and O. J. Sharp. Compiler transformations for high-performance computing. *ACM Comput. Surv.*, 26(4):345–420, Dec. 1994. ISSN 0360-0300.

[6] C. Bastoul. Efficient code generation for automatic parallelization and optimization. In *Proceedings of the Second International Conference on Parallel and Distributed Computing*, ISPDC'03, pages 23–30, 2003. ISBN 0-7695-2069-3.

[7] C. Bastoul. Code generation in the polyhedral model is easier than you think. In *Proceedings of the 13th International Conference on Parallel Architectures and Compilation Techniques*, PACT '04, pages 7–16, 2004. ISBN 0-7695-2229-7.

[8] Y. Ben-Asher and G. Haber. Parallel solutions of simple indexed recurrence equations. *IEEE Trans. Parallel Distrib. Syst.*, 12(1):22–37, Jan. 2001. ISSN 1045-9219.

[9] R. S. Bird. An introduction to the theory of lists. In *Proceedings of the NATO Advanced Study Institute on Logic of Programming and Calculi of Discrete Design*, pages 5–42, 1987. ISBN 0-387-18003-6.

[10] G. E. Blelloch. Prefix sums and their applications. 1990.

[11] W. Blume, R. Doallo, R. Eigenmann, J. Grout, J. Hoeflinger, T. Lawrence, J. Lee, D. Padua, Y. Paek, B. Pottenger, L. Rauchwerger, and P. Tu. Parallel programming with polaris. *Computer*, 29(12):78–82, Dec. 1996. ISSN 0018-9162.

[12] W.-N. Chin, A. Takano, and Z. Hu. Parallelization via context preservatio. In *Proceedings of the 1998 International Conference on Computer Languages*, ICCL '98, pages 153–, 1998.

[13] G. Contreras and M. Martonosi. Characterizing and improving the performance of intel threading building blocks. In *4th International Symposium on Workload Characterization (IISWC 2008), Seattle, Washington, USA, September 14-16, 2008*, pages 57–66, 2008.

[14] G. Fedyukovich and R. Bodık. Approaching symbolic parallelization by synthesis of recurrence decompositions. In *5th Workshop on Synthesis (SYNT)*, 2016. URL `http://formal.epfl.ch/synt/2016/papers/paper10.pdf`.

[15] A. L. Fisher and A. M. Ghuloum. Parallelizing complex scans and reductions. In *Proceedings of the ACM SIGPLAN 1994 Conference on Programming Language Design and Implementation*, PLDI '94, pages 135–146, 1994. ISBN 0-89791-662-X.

[16] A. Geser and S. Gorlatch. Parallelizing functional programs by generalization. In *Proceedings of the 6th International Joint Conference on Algebraic and Logic Programming*, ALP '97-HOA '97, pages 46–60, 1997. ISBN 3-540-63459-2.

[17] J. Gibbons. The third homomorphism theorem. *J. Funct. Program.*, 6(4):657–665, 1996.

[18] S. Gorlatch. Systematic extraction and implementation of divide-and-conquer parallelism. In *Proceedings of the 8th International Symposium on Programming Languages: Implementations, Logics, and Programs*, PLILP '96, pages 274–288, 1996. ISBN 3-540-61756-6.

[19] S. Gorlatch. Extracting and implementing list homomorphisms in parallel program development. *Sci. Comput. Program.*, 33(1):1–27, Jan. 1999. ISSN 0167-6423.

[20] W. D. Hillis and G. L. Steele Jr. Data parallel algorithms. *Communications of the ACM*, 29(12):1170–1183, 1986.

[21] H. Hwansoo and T. Chau-Wen. A comparison of parallelization techniques for irregular reductions. In *Parallel and Distributed Processing Symposium., Proceedings 15th International*, pages 8 pp.–, 2001.

[22] S. Kamil, A. Cheung, S. Itzhaky, and A. Solar-Lezama. Verified lifting of stencil computations. In *Proceedings of the 37th ACM SIGPLAN Conference on Programming Language Design and Implementation*, PLDI '16, pages 711–726, 2016. ISBN 978-1-4503-4261-2.

[23] A. Kejariwal, P. D'Alberto, A. Nicolau, and C. D. Polychronopoulos. A geometric approach for partitioning n-dimensional non-rectangular iteration spaces. In *Proceedings of the 17th International Conference on Languages and Compilers for High Performance Computing*, LCPC'04, pages 102–116, 2005. ISBN 3-540-28009-X, 978-3-540-28009-5.

[24] R. A. Kelsey. A correspondence between continuation passing style and static single assignment form. In *Papers from the 1995 ACM SIGPLAN Workshop on Intermediate Representations*, IR '95, pages 13–22, 1995. ISBN 0-89791-754-5.

[25] R. E. Ladner and M. J. Fischer. Parallel prefix computation. *Journal of the ACM (JACM)*, 27(4):831–838, 1980.

[26] K. R. M. Leino. Dafny: An automatic program verifier for functional correctness. In *Logic for Programming, Artificial Intelligence, and Reasoning (LPAR)*, pages 348–370, 2010.

[27] A. Maaz Bin Safeer and A. Cheung. Leveraging parallel data processing frameworks with verified lifting. In *5th Workshop on Synthesis (SYNT)*, 2016.

[28] C. Marché. Normalized rewriting: an alternative to rewriting modulo a set of equations. *Journal of Symbolic Computation*, 21(3):253–288, 1996.

[29] C. Marché and X. Urbain. *Termination of associative-commutative rewriting by dependency pairs*, pages 241–255. Springer Berlin Heidelberg, Berlin, Heidelberg, 1998.

[30] A. Morihata. A short cut to parallelization theorems. In *ACM SIGPLAN International Conference on Functional Programming, ICFP'13, Boston, MA, USA - September 25 - 27, 2013*, pages 245–256, 2013.

[31] A. Morihata and K. Matsuzaki. Automatic parallelization of recursive functions using quantifier elimination. In *Functional and Logic Programming, 10th International Symposium, FLOPS 2010, Sendai, Japan, April 19-21, 2010. Proceedings*, pages 321–336, 2010.

[32] A. Morihata, K. Matsuzaki, Z. Hu, and M. Takeichi. The third homomorphism theorem on trees: downward & upward lead to divide-and-conquer. In *Proceedings of the 36th ACM SIGPLAN-SIGACT Symposium on Principles of Programming Languages, POPL 2009, Savannah, GA, USA, January 21-23, 2009*, pages 177–185, 2009.

[33] K. Morita, A. Morihata, K. Matsuzaki, Z. Hu, and M. Takeichi. Automatic inversion generates divide-and-conquer parallel programs. In *Proceedings of the 28th ACM SIGPLAN Conference on Programming Language Design and Implementation*, PLDI '07, pages 146–155, 2007.

[34] K. Morita, A. Morihata, K. Matsuzaki, Z. Hu, and M. Takeichi. Automatic inversion generates divide-and-conquer parallel programs. In *Proceedings of the 28th ACM SIGPLAN Conference on Programming Language Design and Implementation*, PLDI '07, pages 146–155, 2007. ISBN 978-1-59593-633-2.

[35] P. Narendran and M. Rusinowitch. *Any ground associative-commutative theory has a finite canonical system*, pages 423–434. Springer Berlin Heidelberg, Berlin, Heidelberg, 1991.

[36] G. C. Necula, S. McPeak, S. P. Rahul, and W. Weimer. *CIL: Intermediate Language and Tools for Analysis and Transformation of C Programs*. 2002.

[37] D. A. Padua and M. J. Wolfe. Advanced compiler optimizations for supercomputers. *Commun. ACM*, 29(12):1184–1201, Dec. 1986. ISSN 0001-0782.

[38] C. H. Papadimitriou and M. Sipser. Communication complexity. *J. Comput. Syst. Sci.*, 28(2):260–269, 1984.

[39] C. Pheatt. Intel® threading building blocks. *Journal of Computing Sciences in Colleges*, 23(4):298–298, 2008.

[40] K. Pingali, D. Nguyen, M. Kulkarni, M. Burtscher, M. A. Hassaan, R. Kaleem, T.-H. Lee, A. Lenharth, R. Manevich, M. Méndez-Lojo, D. Prountzos, and X. Sui. The tao of parallelism in algorithms. *SIGPLAN Not.*, 46(6):12–25, June 2011. ISSN 0362-1340.

[41] C. Radoi, S. J. Fink, R. Rabbah, and M. Sridharan. Translating imperative code to mapreduce. In *Proceedings of the 2014 ACM International Conference on Object Oriented Programming Systems Languages & Applications*, OOPSLA '14, pages 909–927, 2014. ISBN 978-1-4503-2585-1.

[42] V. Raychev, M. Musuvathi, and T. Mytkowicz. Parallelizing user-defined aggregations using symbolic execution. In *Proceedings of the 25th Symposium on Operating Systems Principles*, SOSP '15, pages 153–167, 2015. ISBN 978-1-4503-3834-9.

[43] S. Sato and H. Iwasaki. Automatic parallelization via matrix multiplication. *SIGPLAN Not.*, 46(6):470–479, June 2011. ISSN 0362-1340.

[44] C. Smith and A. Albarghouthi. Mapreduce program synthesis. *SIGPLAN Not.*, 51(6):326–340, June 2016. ISSN 0362-1340.

[45] A. Solar-Lezama, G. Arnold, L. Tancau, R. Bodik, V. Saraswat, and S. Seshia. Sketching stencils. *SIGPLAN Not.*, 42(6):167–178, June 2007. ISSN 0362-1340.

[46] A. Solar-Lezama, C. G. Jones, and R. Bodik. Sketching concurrent data structures. In *Proceedings of the 29th ACM SIGPLAN Conference on Programming Language Design and Implementation*, PLDI '08, pages 136–148, 2008. ISBN 978-1-59593-860-2.

[47] E. Torlak and R. Bodík. Growing solver-aided languages with rosette. In *ACM Symposium on New Ideas in Programming and Reflections on Software, Onward! 2013, part of SPLASH '13, Indianapolis, IN, USA, October 26-31, 2013*, pages 135–152, 2013.

[48] N. Vasilache, C. Bastoul, and A. Cohen. Polyhedral code generation in the real world. In *Proceedings of the 15th International Conference on Compiler Construction*, CC'06, pages 185–201, 2006. ISBN 3-540-33050-X, 978-3-540-33050-9.

[49] P. Černý, T. A. Henzinger, A. Radhakrishna, L. Ryzhyk, and T. Tarrach. Regression-free synthesis for concurrency. In *Proceedings of the 16th International Conference on Computer Aided Verification - Volume 8559*, pages 568–584, 2014. ISBN 978-3-319-08866-2.

[50] M. Vechev, E. Yahav, and G. Yorsh. Abstraction-guided synthesis of synchronization. In *Proceedings of the 37th Annual ACM SIGPLAN-SIGACT Symposium on Principles of Programming Languages*, POPL '10, pages 327–338, 2010. ISBN 978-1-60558-479-9.

[51] R. Wilson, R. French, C. Wilson, S. Amarasinghe, J. Anderson, S. Tjiang, S. Liao, C. Tseng, M. Hall, M. Lam, and J. Hennessy. The suif compiler system: A parallelizing and optimizing research compiler. Technical report, Stanford, CA, USA, 1994.

[52] L. D. Zuck, A. Pnueli, Y. Fang, B. Goldberg, and Y. Hu. Translation and run-time validation of optimized code. *Electr. Notes Theor. Comput. Sci.*, 70(4):179–200, 2002.

## A. Conversion to a system of equations

We show that we can use a simple procedure to convert a loop body given in the input langauge described in 3 without loops into a system of equation. We proceed statement by statement, updating expressions on the right-hand side of an equation system and merging branches of conditionnals ($\varphi$ functions in static single assignments). First, let us remark that we can convert all if-then statements to if-then-else statements by adding an empty else statement.

We start with a system where each varaible in SVar is assigned to itself : $E = \langle s_0 = s_0, \ldots, s_n = s_n \rangle$, and update it by visiting each statement following the control flow graph edges (we do not have loops). We transform $E$ into the new system $E'$, depending on what statement we see:

- an assignment $s_i$ = Exp(SVar, IVar). The expression of $s_i$ in $E$ is updated by Exp(SVar, IVar)$[s_j \leftarrow exp_j \in E]$, $E'$ is $E$ with $s_i$ = Exp once all the variables appearing in Exp have been replaced by their expression in $E$.

- a conditional on expression $c$ : we apply the procedure recursively, each branch is converted to a system of equations $E_{if}$ and $E_{else}$. The systems are merged into :

$$E_{merged} = \langle \ldots s_i = (c \; ? \; exp_i^{if} \; : \; exp_i^{else}), \ldots \rangle$$

where $exp_i^{if}$ is the expression of $s_i$ in $E_{if}$ and $exp_i^{else}$ is the expression of $s_i$ in $E_{else}$.
We update $E$ to be:

$$E' = \langle \ldots exp_i^{merged}[s_j \leftarrow exp_j \in E] \ldots \rangle$$

with the susbtitution of variables in the expressions of the merged system by their expression in $E$.

And then we proceed to the next statement.

This procedure would yield systems with expressions of non optimal size, but this supports our claim that we can convert any input loop body into a system of equations.

## B. Proof Sketch of Proposition 4.4

If the theorem of parallelization with the weak right inverse in [34] holds and the weak-right inverse returns a list of constant length, then there is a solution to the synthesis problem. The theorem states that the join $\odot$ for the loop body function $f_E$ can be built using the following construct :

$$a \odot b = f_E(f_E^0 \; a \bullet f_E^0 \; b)$$

where $f_E^0$ is the weak-right inverse of function $f_E$. The recursive application of $f_E$ on the two concatenated lists $f_E^0 a$ and $f_E^0 b$ will then yield a vector of expressions, in which all the occurences of the state variables in the expresssions $Exp_i$ will contain an expression containing variables from inputs $a$ and $b$, whereas the last input read contains only variables from the right input $b$. The solution given by this procedure is therefore a completion of the sketch for the problem.

## C. Proof of Proposition 5.2

Consider the figure below that illustrates the parallel computation of a $\odot$-homomorphic function $f'$ over the list $x = [x_1, \ldots, x_n]$:

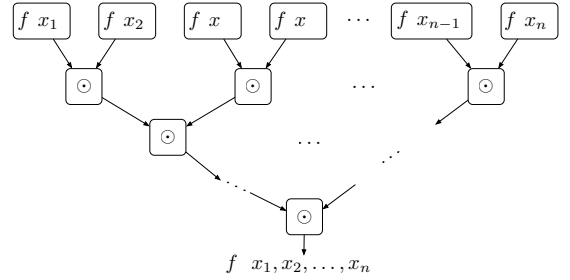

One can use the recurrence

$$J(n) = 2J(n/2) + t_j(n)$$

to compute the time complexity of performing the join operations, where $n$ corresponds to the size of input handled by join, and $ct_j(n)$ captures the cost of performing one join of size $n$. The entire cost of computing $f'(x)$ in parallel (as illustrated in the diagram above) will then be obtained by adding to $J(n)$ the cost of computing $f'(x_i)$ for each $x_i$, which is constant for each $f'(x_i)$ and therefore linear overall. Therefore the cost of computing $f'$ in parallel is in total:

$$T_p(n) = J(n) + cn$$

To have an efficient parallel implementation (with constantly many processors[2]), we need to our parallel asymptotic time complexity for $f'$ not to be larger than our asymptotic sequential time complexity for $f$, that is $T_p(n) \in O(n)$.

The Master theorem then suggests that for $T_p(n) \in O(n)$ to hold, we have to have $t_j(n) \in o(n)$. That is, the cost of join should be sub-linear.

## D. Proof of Theorem 6.1

A more detailed version of the theorem statement is:

If $f : \mathcal{S} \to Sc$ is both leftwards and rightwards linear time, then there exists a tuple of functions $\langle g_1, \ldots, g_k \rangle$ for a constant $k$, where $g_i : \mathcal{S} \to Sc$ ($1 \leq i \leq k$), and an operator $\odot$ such that $f \bowtie g_1 \bowtie \ldots \bowtie g_k$ is $\odot$-homomorphic.

By the third homomorphism theorem, we know that $f$ is $\odot$-homomorphic for some $\odot$. All that remains to show is that there exists a constant-time $\odot$.

If $f$ is rightwards linear time then

$$f(y \bullet a) = f(y) \otimes a$$

such that $\otimes$ is constant time computable. Similarly, if $f$ is leftwards linear time then

$$f(a \bullet y) = a \oplus f(y)$$

---
[2] This is just to clarify that the discussion is avoiding common scenarios in the PRAM model that the number of processors available can be a function of the input size, such as $\log(n)$.

such that ⊕ is constant time computable.

We prove the claim by double induction on the sizes of the split sequences. Consider the base case $y = [\,]$: if $f([\,])$ exists, then since $f(x) \odot f([\,]) = f(x)$, it is obviously constant time computable (there is nothing to be done if $f(x)$ is available). If $f([\,])$ does not exist, the $y = a$ is the base case and $f(x) \odot f([a]) = f(x) \otimes a$, and by the assumption above, it is constant-time computable.

The symmetric left argument proves the base cases $x = [\,]$ and $x = a$ for $x$.

Let us assume $f(x) \odot f(y)$ is constant-time computable for $|x|, |y| < n$.

$$f(x \bullet (y \bullet a)) = f(x) \odot f(y \bullet a)$$
$$= f(x \bullet y) \otimes a$$
$$= (f(x) \odot f(y)) \otimes a$$

The last term is constant time computable by induction hypothesis and the assumptions. Therefore, $f(x) \odot f(y \bullet a)$ is constant time computable. So, we have just proven the induction step for $y$. Similarly:

$$f((a \bullet x) \bullet y) = f(a \bullet x) \odot f(y)$$
$$= a \oplus f(x \bullet y)$$
$$= a \oplus (f(x) \odot f(y))$$

The last term is constant time computable by induction hypothesis and the assumptions. Therefore, $f(a \bullet x) \odot f(y)$ is constant time computable. And, now we have proven the induction step for $x$ as well.

### E. Proof of Theorem 6.2

**Lemma E.1** (from [17]). *A function $h : \mathcal{S} \to Sc$ is a homomorphism iff for all sequences $x, y, v,$ and $w$, we have*

$$h(x) == h(v) \land h(y) = h(w) \implies h(x \bullet y) = h(v \bullet w)$$

Assume a rightwards $f : \mathcal{S} \to Sc$ is not homomorphic, and is extended by a function $g$ into a ⊙-homomorphic function $f \ltimes g$.

According to the third homomorphism theorem [17], if a function is homomorphic, then it is both leftwards and rightwards. We will use the fact that $f$ is rightwards, and that $h = f \ltimes g$ is both leftwards and rightwards in the argument below.

Assume that for four arbitrary sequences $x$, $y$, $v$ and $w$, we have

$$f(x) = f(v) \land h(y) = h(w)$$

the latter being logically equivalent to $f(y) = f(w) \land g(y) = g(w)$. Then, we have:

$$(f(x \bullet y), g(x \bullet y))$$
$$= h(x \bullet y) \quad \text{by definition}$$
$$= h(x \bullet w) \quad \text{by Lemma E.1}$$
$$= (f(x \bullet w), g(x \bullet w)) \quad \text{by definition}$$
$$= (f(v \bullet w), g(x \bullet w)) \quad \text{by } f \text{ being rightwards}$$

We have just argued for a variation of Lemma E.1:

$$f(x) = f(v) \land f(y) = f(w) \land g(y) = g(w)$$
$$\implies f(x \bullet y) = f(v \bullet w)$$

which effectively says that the value of $f(x \bullet y)$ only depends on values of $f(x)$, $f(y)$, and $g(y)$; and, specifically not $g(x)$. Now, let us consider the join operator ⊙ for $f \ltimes g$. Since the return values of $f$ and $g$ are considered to be scalars, we have:

$$(f(x), g(x)) \odot (f(y), g(y)) =$$
$$(exp(f(x), g(x), f(y), g(y)), exp'(f(x), g(x), f(y), g(y)))$$

The argument above indicates that the dependency of $exp$ on $g(x)$ must be non-existent or superficial. In other words, the join can be restructured in the following form:

$$(f(x), g(x)) \circledast (f(y), g(y)) =$$
$$(exp(f(x), f(y), g(y)), exp'(f(x), g(x), f(y), g(y)))$$

which brings us to the proof of Theorem 6.2.

### F. Proof of Theorem 6.3

Let $f$ be a rightwards function (which can be a tuple of many functions) that is sequentially defined as

$$f(y \bullet a) = f(y) \otimes a$$

Let us assume that there is a function $g$ (which can be a tuple of many functions) such that $f \ltimes g$ is ⊙-homomorphic (constant) for some ⊙. By Theorem 6.2, then we know that there exist ⊛, exp, and $exp'$ such that:

$$f \ltimes g(x \bullet y) = f \ltimes g(x) \circledast f \ltimes g(y) =$$
$$(\mathsf{exp}(f(x), f(y), g(y)), \mathsf{exp}'(f(x), g(x), f(y), g(y)))$$

The assumptions of the theorem about the existence of the algebraic rules implies that there is a sequence of rules that if applied step by step, they would transform the lefthand side of the above equation to the righthand side. Therefore, we can assume that the exist a run of *normalize* that does exactly that, and returns the righthand side.

Since $f \ltimes g$ is a homomorphism, by the third homomorphism theorem, it is rightwards. Therefore, there exists ⊕ such that:

$$f \ltimes g(y \bullet a) = f \ltimes g(y) \oplus a$$

this means that $g$ can be computed sequentially along $f$. Note that without this argument, it would not be the case that $g$ corresponds to an accumulator. Once we know it does, it is straightforward to see that if *normalize* returns the righthand side to the *solve*, then *solve* will discover the correct accumulator.

## G.  Proof of Remark 6.3

The rightwards function corresponds to the regular language $L_1 = \{w|\ w[n] = 1\}$. It is clear that a DFA for this language should just be able to count up to $n$ using $n$ states, and then check one letter.

The leftwards function corresponds to the regular language $L_2 = \{w|\ \textit{n-th bit from the end is}\ 1\}$. This language cannot be as efficiently as $L_1$. The DFA for $L_2$ needs to record all the last $n$ bits in its state, so that with every new input letter, it can update the window of size $n$, being prepared for the unknown end of the string to come at any time. Once the string is finished, then it has to look at the first letter in the $n$-letter window that it has memorized to see if it is a 0 or a 1.

To prove that one cannot do better than the above, a simple distinguishability argument will do. Let $u$ denote the string consisting of the last $n$ letters of the input string $w$. Consider now that the input string is extended by a sequence of strings $v_0, v_2, \ldots v_{n-1}$ where each $v_i$ has length $i$. For each $v_i$, for the DFA to correctly accept/reject $wv_i$, it has to be able to distinguish the cases $u[i] = 0$ and $u[i] = 1$. So, if the DFA is in state $q$ after reading letter $u[i-1]$, it has to go to two different states for the two different values of $u[i]$. Therefore, the DFA needs at least $2^n$ different states to distinguish all cases that potentially would lead to different acceptance/rejection results depending on the extensions of the string.